# Default Supply Auctions in Electricity Markets: Challenges and Proposals

Juan Ignacio Peña[a] and Rosa Rodriguez[b]

**Abstract**

This paper studies premiums got by winning bidders in default supply auctions, and speculation and hedging activities in power derivatives markets in dates near auctions. Data includes fifty-six auction prices from 2007 to 2013, those of CESUR in the Spanish OMEL electricity market, and those of Basic Generation Service auctions (PJM-BGS) in New Jersey's PJM market. Winning bidders got an average ex-post yearly forward premium of 7% (CESUR) and 38% (PJM-BGS). The premium using an index of futures prices is 1.08% (CESUR) and 24% (PJM-BGS). Ex-post forward premium is negatively related to the number of bidders and spot price volatility. In CESUR, hedging-driven trading in power derivatives markets predominates around auction dates, but in PJM-BGS, speculation-driven trading prevails. The policy recommendation to market regulators and administrators is that they should gauge consumers' price risk aversion before introducing alternative methods to default supply auctions for the computation of the part of cost of energy of the electricity bill of customers whose contracted capacity is small and are not served by other suppliers.

[a] Corresponding author. Universidad Carlos III de Madrid, Department of Business Administration, c/ Madrid 126, 28903 Getafe (Madrid, Spain). ypenya@eco.uc3m.es; [b] Universidad Carlos III de Madrid, Department of Business Administration, c/ Madrid 126, 28903 Getafe (Madrid, Spain). rosa.rodriguez@uc3m.es.

# Default Supply Auctions in Electricity Markets: Challenges and Proposals

## Abstract

This paper studies premiums got by winning bidders in default supply auctions, and speculation and hedging activities in power derivatives markets in dates near auctions. Data includes fifty-six auction prices from 2007 to 2013, those of CESUR in the Spanish OMEL electricity market, and those of Basic Generation Service auctions (PJM-BGS) in New Jersey's PJM market. Winning bidders got an average ex-post yearly forward premium of 7% (CESUR) and 38% (PJM-BGS). The premium using an index of futures prices is 1.08% (CESUR) and 24% (PJM-BGS). Ex-post forward premium is negatively related to the number of bidders and spot price volatility. In CESUR, hedging-driven trading in power derivatives markets predominates around auction dates, but in PJM-BGS, speculation-driven trading prevails. The policy recommendation to market regulators and administrators is that they should gauge consumers' price risk aversion before introducing alternative methods to default supply auctions for the computation of the part of cost of energy of the electricity bill of customers whose contracted capacity is small and are not served by other suppliers.

1. Introduction

Liberalization processes of electricity markets around the world face many challenges, such as how to supply electricity to different customers at prices consistent with market circumstances. Auction mechanisms have played a salient role in many countries in the effort to match supply and demand as an alternative to other pricing systems, although the theoretical conclusions and empirical evidence are ambiguous, Newbery and McDaniel (2003). In deregulated markets, auctions are a mechanism applied in many countries to supply electricity to customers whose contracted capacity is small and are not served by other suppliers. Providers of last resort (POLR), designated by the corresponding public utility commission, must get electricity from somewhere and must sell energy to those customers. Default Supply Auction (DSA) is a method to supply electricity to POLR, Loxley and Salant (2004). In these auctions, POLRs buy electricity forward contracts from winning bidders (WB) at prices determined by the specific auction mechanisms (e.g. sealed bid, ascending auctions, descending clock auction (DCA)[1]). Market regulators use DSA-based prices for computing the variable factor of the cost of energy part. When choosing DSA to give electricity to the POLRs, market administrators assume at least two hypotheses: (1) DSA provides efficient generation resources at competitive prices[2], and (2) DSA gives agents incentives to engage in hedging activities, using power derivatives[3]. This paper studies the extent to which these two hypotheses are consistent with empirical evidence from actual DSA experiences in Spain (CESUR auction) and in the State of New Jersey (PJM-BGS auction) in 2007-2013. Retail electricity prices contain two elements: (i) the cost of supplying electricity and (ii) the "government wedge" (taxes, levies, and other charges to finance public policies). In this paper, we focus on (i), "the cost of energy", composed of two factors: a fixed factor related to contracted capacity and a variable factor related to electricity prices. We analyse DSA-based prices that regulators use when setting the amount to be charged to consumer corresponding to the variable factor.

---

[1] In ascending auctions, the auctioneer begins with a low asking price for the product being acquired, which is increased by bids from participants. Price and allocation are determined in an open competition among the bidders. The bidders willing to pay the most win. In DCAs, in each round the Auctioneer announces a price for the product being acquired. Bidders bid for the right to provide the quantity of the product they wish to supply at the price announced by the Auctioneer. Bidders decide what quantity of the product they wish to offer to provide in a particular round of the auction. Following the end of a round, the Auctioneer adds up all bids received at the price for that round. If the total quantity of the product bid is greater than the quantity to be acquired, the Auctioneer announces a lower price for the following round. Bidders then decide how much to offer to supply at the new, lower price. The quantity of the product that a bidder offers to supply in the next round can be the same as or smaller, but not larger, than it offered in a previous round. A bidder must submit bids in every round, and cannot re-enter the auction once it abstains from bidding in a round. When the total quantity bid by all bidders matches the total quantity sought by the Auctioneer, the auction closes. The winners are the bidders in the last successful round of the auction.

[2] According to New Jersey Board of Public Utilities (BPU), the auction is designed to procure supply for PJM-BGS customers "at a cost consistent with market conditions". More details of PJM-BGS auctions can be found at http://www.BGS-auction.com/BGS.auction.overview.asp. The Royal Decree 1634/2006 regulating CESUR auction states: "the goal is to adapt tariffs (auction prices) to market prices". CESUR auctions, (see Order ITC/400/2007) help in the pricing of the energy component included in tariffs charged to final consumer. They also intend to prevent further tariff deficits.

We present an empirical framework showing that auction-related factors (e.g. number of bidders) and market factors (e.g. spot price volatility) help in explaining the ex-post forward premium in CESUR and PJM-BGS. Besides that, we show that hedging and speculative activity in power derivatives markets increases in dates near the auctions. An empirical simulation in the period 2007-2013 suggest that consumer's aversion to price volatility plays a key role when choosing between alternative methods for the computation of the variable factor of the cost of energy.

This paper contributes to the literature on the empirical analysis of ex-post forward premium in CESUR auctions by completing partial results in Federico and Vives (2008), Cartea and Villaplana (2012), Fabra and Fabra Utray (2012) and Capitán Herráiz (2014). Ours is the first paper analysing the full set of auctions and documenting a significant relation between ex-post forward premium and the number of participants in the auction. We also contribute to the study of the results of PJM-BGS auctions, extending results in Loxley and Salant (2004), Lacasse and Wininger (2007) and Castro, Negrete-Pincetic and Gross (2008). Different from their papers, we present an empirical model explaining ex-post forward premium and trading activity in power derivatives markets. Another novel contribution is that, in both markets, we compute premium during delivery periods, by comparing auction prices against (i) spot (day-ahead) prices and (ii) the Forward Market Price Index (FMPI) containing prices of forward contracts matching the service period. Finally, this paper contributes to the literature on the extent to which default supply auctions in electricity markets obtain electricity prices consistent with market conditions (Maurer and Barroso, 2011). The empirical findings in this paper give suggestions to be considered in the current discussions in energy policy about methods for supplying electricity to customers whose contracted capacity is small and are not served by other suppliers.

---

[3] According to New Jersey Board of Public Utilities (BPU), auctions provide an opportunity for energy trading and marketing companies to provide PJM-BGS supply. One key goal of CESUR auctions is "encourage forward contracting", CNE (2008).

The rest of this paper is organized as follows. Section 2 discusses the main characteristics of CESUR and PJM-BGS auctions. Section 3 presents the methodology of measures for hedging and speculation in derivatives markets. We present the data in Section 4. The empirical analysis is in Section 5. Section 6 concludes with policy recommendations.

## 2. Default Supply Auctions: CESUR and PJM-BGS

### 2.1 CESUR

Ministerial Order ITC/400/2007 implemented a quarterly auction (CESUR auctions); to support the calculation of the energy price to be passed through[4] to regulated consumers[5]. The mechanism is as follows. The government (Secretaría General de la Energía, SGE) announces the amount of energy to be auctioned, first price and auction dates. SGE also announces delivery periods. There are five POLR, designated by the government. The government sets their share as buyers in CESUR auctions. The firms (shares) are ENDESA (35%), IBERDROLA (35%), EDP (12%), FENOSA (11%), HIDROCANTÁBRICO (4%) and VIESGO (3%)[6]. Winning bidders (WB) sell forward contracts to POLRs. The Government appointed the National Energy Commission as the trustee of the auctions (CNE, Comisión Nacional de Energía, later subsumed into the CNMC, Comisión Nacional de los Mercados y la Competencia, from October 2013). CNE contracted an independent consulting firm to conduct the first five auctions, which started in June 2007. From the sixth auction onward until December 2013, the managing body responsible for organizing and managing the auctions has been OMEL, the electricity market operator. The 25th CESUR auction (December 19, 2013) produced a final price deemed "too high" and the Spanish government annulled the auction on allegations of "manipulations"[7]. Since then, they suspended CESUR auctions[8].

---

[4] Final consumer electricity prices contain two elements: (i) cost of supplying electricity (including the price of energy based on CESUR prices in 2007-2013 plus the regulated cost of providing network services) and (ii) national "government wedge". This wedge includes taxes, levies, and other charges to finance public policies such as feed-in tariff support to renewables and payments of interests to investors in securitized 'tariff debt' traded in international financial markets. Spain's "government wedge" is the second highest in Europe (after Germany) and is the main reason for the rise in retail electricity prices since 2008; see Robinson (2015). It accounts for approximately 60% of final electricity prices.

[5] Households and small firms connected in low tension (< 1kV) and contracted load lower or equal than 10 kW. 22 million consumers in 2015.

[6] After the 5th auction, shares changed as follows: ENDESA 29%, IBERDROLA 40%, EDP 12%, FENOSA 6%, HIDROCANTABRICO 12%, VIESGO 1%.

[7] This is a controversial issue. See the extensive report by the market regulator CNMC (2014) which points out to some "atypical" circumstances. For instance, the auction closed after seven rounds. In previous cases, the minimum figure was twelve rounds.

[8] From July 2007 until March 2014, the price of energy part included in retail prices was referenced to CESUR prices. Since April 2014 to the present, the Spanish Government adopted a new system based on PVPC (*Precio Voluntario para el Pequeño Consumidor,* Volunteer prices for small consumers) tariffs in which the price of energy components is calculated using hourly prices of the wholesale electricity market. Therefore, small consumers affiliated to PVPC are exposed to daily fluctuations in prices, that is, to market price risk. Such exposure causes growing concern both to consumers and to government officials. Small consumers also can sign "free-market" contracts with commercial suppliers.

In the 24 CESUR auctions they offered 28 different products. The quantities auctioned are always smaller than the amount needed to fulfill regulated consumers' needs. In most cases, auctions included three-month base-load contracts, amounting to an average of 3,500 MW, or less than 30 percent of expected needs. An average of thirty domestic and international allowed bidders take part in the bidding and contracts are awarded to an average of fifteen WB, including retailers, generators, and marketers. All CESUR auctions are simultaneous descending clock auctions. On average, auctions closed after twenty-three rounds. Cash flows between WB and POLR resulting from CESUR auctions are computed each hour during the delivery period by differences between CESUR prices and spot (wholesale) market prices.

**2.2 PJM-BGS**

Since 2002 to the present, four designated POLR: Public Service Electric & Gas Company (PSE&G), Atlantic City Electric Company (ACE), Jersey Central Power & Light Company (JCP&L), and Rockland Electric Company (RECO) serve PJM's Basic Generation Service (PJM-BGS) customers through auctions held in February. Each POLR serves a specific geographic area (ACE, PSEG, JCPL, RECO) within the overall PJM system[9]. Each area has its specific spot (day-ahead) price. BGS refers to the service of customers who are not served by a third-party supplier or competitive retailer. Two auctions are held concurrently, one for larger customers (BGS-CIEP) and one for smaller commercial and residential customers (BGS-RSCP, BGS-FP). In this paper, we concentrate on the latter because its final customers are like CESUR's. We call it PJM-BGS. An average of nineteen allowed bidders takes part in the bidding and contracts are awarded to an average of seven (PSEG), six (JCPL), four (ACE) and one (RECO) winning bidders. Considering all zones together, the average number of different winning bidders is ten [10]. On average, auctions closed after twenty-two rounds. They fulfil customer needs through rolling three-year contracts. Each year, one-third of the contracts run out and are replaced with new three-year contracts. Therefore, prices paid by final consumers are averages of prices from three auctions. Since auction takes place in February in year *t*, WB must sell electricity during thirty-six months from June 1st (year *t*) through May 31st (year *t*+3). In contrast with CESUR, the products in PJM-BGS are full requirements. Thus, besides energy costs, auction prices contain distribution and capacity costs, the cost of ancillary services and market price risk. To make results from CESUR and PJM-BGS comparable, we take off from PJM-BGS auction prices all observable factors (distribution and capacity costs and the cost of ancillary services). A WB supplies a percentage of a POLR's load, whatever the load may be. therefore, a WB who wins 10% of a POLR's load provides all services necessary to serve 10% of that POLR's load. Note that WB bear risks associated with load level. So, WB are exposed to price and quantity (volumetric) risk. Winning bidders receive an all-in payment based on the auction price for a POLR. In summer and winter, each WB receives auction price times a POLR-specific season factor for every kWh of load served for that POLR. The Summer payment factor is around 1.2 and winter payment factor is around 0.9. BGS-RSCP customers pay retail rates in which the energy part derives from BGS auction prices. More details on PJM-BGS auctions are in Lacasse and Wininger (2007).

---

[9] The four utilities decided to sell their generation assets or transfer them to affiliates, thus becoming electric distribution companies (EDC). An EDC is a "wires only" company. This means that the only assets it owns are wires (a "natural monopoly") and the firm is within the ambit of the market regulator.

---

[10] A bidder may get auction tranches in different zones. For instance, in the 2013 auction NextEra Energy Power Marketing LLC obtained four tranches in PSEG, two tranches in JCPL, three tranches in ACE and one tranche in RECO. WBs become BGS suppliers and provide full-requirements service for final customers. Full-requirements service includes capacity, energy, ancillary services and transmission, and any other service as PJM may require. WBs assume migration risk and must satisfy the state's renewable energy portfolio standard.

In summary, the key differences between CESUR and PJM-BGS prices are (i) CESUR prices refer to the cost of energy (cost of generation) but PJM-BGS prices contain many other elements (distribution, transmission, ancillary costs) besides the cost of energy, and (ii) CESUR prices refer to a fixed amount of electricity but PJM-BGS prices are full requirements and refer to a part of total load, whatever that load turn out to be.

## 3. Methodology: Measures of hedging and speculation

We are interested in measuring relative activity of hedgers and speculators in power derivatives markets. Key variables to be considered are daily trading volume and open interest, see Leuthold (1983) and Bessembinder and Seguin (1993). For a contract, daily trading volume accounts for its trading activity and reflects movements in speculative trading because this measure includes intra-day positions. Open interest is the number of outstanding contracts at the end of the trading day and captures hedging activity, because, by definition, excludes all intra-day positions[11].

A natural step is to mix both variables to produce a measure of relative trading activity of hedgers and speculators. García, Leuthold, and Zapata (1986) suggest the volume-to-open-interest ratio, defined:

$$R1(t) = \frac{V(t)}{OI(t)} \qquad (1)$$

Notice that R1 is the ratio of a flow variable (V(t)) with respect to a stock variable (OI(t)). The behavior of R1(t) depends on the history of the contract up to day t, besides market activity on day t. For that reason, ap Gwilym, Buckle and Evans (2002) proposed an alternative measure by relating two flow variables, the ratio of volume to absolute change in open interest, defined:

$$R2(t) = \frac{V(t)}{|\Delta OI(t)|} \qquad (2)$$

---

[11] Many studies support the underlying assumption that hedgers tend to hold their positions longer than speculators, see for instance Ederington and Lee (2002).

In which ΔOI(t) = OI(t) – OI(t-1). So, R2(t) depends only on market activity on day t. Furthermore, R2(t) discriminates between short-term speculation (day trades), which impact on V(t) but do not impact on ΔOI(t) , and the newly taken hedging positions (held overnight) which impact equally on V(t) and on ΔOI(t) . The higher the relative importance of speculative (hedging) demand is, the higher (lower) the value of R1(t) and R2(t). This suggest a positive correlation between R1(t) and R2(t)[12].

## 4. Data

In this section we present data sources and database providers. The sample period is from January 1, 2007 to December 31, 2013. Auction data comes from fifty-six auction prices got from twenty-four CESUR auctions with twenty-eight products and prices and seven PJM-BGS auctions in four zones, yielding twenty-eight prices. When aggregating the data in yearly figures, we get seven observations in CESUR and twenty-eight in PJM-BGS, totalling thirty-five prices. The sources are the web pages of OMEL and PJM-BGS[13] Day-ahead prices in OMEL are for Spain and in PJM include zones ACE, JCPL, PSEG and RECO[14]. The information about the costs (capacity, transmission, and ancillary services) in PJM we take off from PJM-BGS auction prices to make them comparable with CESUR prices can be found in Monitor Analytics Reports[15]. OMEL FMPI price is the price of the quarterly baseload forward contract (Q1 or Q2) matching the service period. The operator of the Iberian power derivatives market OMIP[16] provided data. PJM-BGS FMPI is based on a strip of futures contracts corresponding to PJM Western Hub Day-Ahead Peak Calendar-Month 5 MW (J4) provided by CME Group[17].

---

[12] Lucia and Pardo (2010) discuss the relative merits of each measure and propose alternative measures.
[13] http://www.subastascesur.omie.es/subastas-cesur/resultados. http://www.bgs-auction.com/.
[14] Data sources are www.omel.es and www.pjm.com.
[15] http://www.monitoringanalytics.com/reports/PJM_State_of_the_Market/2017.shtml contains the data, specifically Vol. 1, 2017, Table 1-9; Total price per MWh by category: Calendar Years 1999 through 2016.

## 5. Empirical Analysis

In the following sections, we study to what extent DSA provides efficient generation resources at competitive prices, and whether DSA gives incentives to agents to engage in hedging activities in power derivatives markets, trading futures contracts.

### 5.1. Premium

In this section, we study whether premiums on CESUR and PJM-BGS auctions are consistent with the purported goal of providing efficient generation resources at competitive prices. Although market regulators do not give a clear definition of competitive prices, one possibility is to interpret this definition implies auction prices should be close to actual wholesale market prices during delivery periods. The difference between the auction price and observed wholesale price during the delivery period is known as the ex-post forward premium, used in many studies. Therefore, if auction price is competitive in this sense, the ex-post forward premium should be small. Another possibility is to assume that auction prices depend on bidder's expectations of the upcoming delivery period. A readily available measure of the expected spot price is the futures price. So, the price of a futures contract matching the service period might be another benchmark[18]. If the difference between the auction price and the futures price is small, auction prices may be competitive. We consider both measures. Winning bidders are taking price risk by selling forward contracts and may want a risk premium. This risk premium should suggest they set auction prices above expected spot prices[19]. The reason is generation costs (for fossil generators) fluctuate daily and spot market prices show these fluctuations, so providing a natural hedge to producers selling electricity in the spot market. But sellers of forward contracts (if they are fossil generators) are exposed to higher volatility in their profits and need a compensation for this situation. Non-fossil generators do not face those risks. Therefore, if fossil generators are predominant, we should expect positive ex-post premiums because sellers need a compensation for assuming price risk[20].

---

[16] www.omip.pt
[17] http://www.cmegroup.com/trading/energy/electricity/pjm-western-hub-peak-calendar-month-day-ahead-lmp-swap-futures.html

To shed light on this point, we compute premium during delivery periods, by comparing auction prices against (i) spot prices [21] (SP) in OMEL and PJM markets during delivery periods and (ii) the Forward Market Price Index (FMPI) containing prices of forward contracts matching the service period, as observed on auction day[22]. Basic facts of CESUR

---

[18] We thank an anonymous referee for this suggestion.

[19] In addition to that, the objective of reducing volatility of final prices serves the interest of risk-averse consumers wishing to avoid the price fluctuations that are usual in electricity spot markets. Presumably, such risk-averse consumers shall pay a moderate risk premium for the security of more stable prices.

[20] This situation contrasts with other derivatives markets in which buyers require this compensation, and in consequence, forward prices tend to be below expected spot prices.

[21] Spot prices in CESUR and PJM are daily averages of (zonal) hourly day-ahead prices. An alternative measure of spot prices could be real-time prices. Theory and empirical evidence in Ito and Reguant (2016) suggest that day-ahead prices are higher than real-time prices in the Iberian market, because of imperfect competition and restricted entry in arbitrage. Therefore, the ex-post premium got by comparing auction prices against day-ahead prices is lower in the Iberian market than the one got using real-time prices. So, our results for this market should be considered a measure of the lower bound of the ex-post forward premium. In the PJM market, Haugom and Ullrich (2012) show that short-term forward prices have converged towards unbiased predictors of the subsequent real-time (spot) prices. Day-ahead prices in OMEL are for Spain and in PJM include zones ACE, JCPL, PSEG and RECO.

[22] CESUR auctions were completed in one day. Since PJM-BGS auctions lasted for three or four days, we use data corresponding to the first day of the auction. Using data for subsequent days or the last day of the auction do not materially affect results.

auctions are in Table 1 (base-load contracts).[23]

[INSERT TABLE 1 HERE]

Columns in Table 1 contain auction number, date, product[24], auction price (CESUR price), average spot price during the delivery period (SP), ex-post forward premium, percentage of ex-post forward premium over CESUR price, average yearly premium, FMPI prices, FMPI premium and percentage FMPI premium over CESUR price. The ex-post forward premium in percentage (the difference between CESUR price and average OMEL spot price during the delivery period divided by CESUR price) ranges from -24.27% to 35.63%. Yearly averages are positive in six out of seven cases and their total average is 7.22%. While winning bidders almost always got a positive yearly premium, consumers supplied by POLRs from 2007 to 2013 paid near €1,000 million over spot market prices[25]. With FMPI, the percentage FMPI premium (defined as the difference between CESUR price and FMPI divided by CESUR price) ranges from –0.34% to 5.69%. Yearly averages are positive, and their average is 1.08%. OMEL offers futures contracts matching the delivery period and auction bidders know the quantity to be supplied before to the actual delivery. Therefore, there is no volumetric risk and winning bidders have at their disposal financial instruments to hedge price risk.

Table 2 reports results on PJM-BGS for each POLR and zone.

---

[23] We analyze baseload contracts only. CESUR introduced Peak load contracts in 2008 but they amount to less than 10% of total capacity auctioned.

[24] For instance, product Q3-07 refers to a contract for delivering electricity from Monday to Sunday, 00:00-24:00, during July, August and September of 2007.

[25] We calculate economic results in millions of Euro by considering capacity offered in each auction, conversion factors, and comparing CESUR prices with wholesale spot prices. The conversion factor implies that each MW of Capacity is equivalent to 2,200 MWh. Detailed results are available on request.

[INSERT TABLE 2 HERE]

Auction BGS-FP price is the auction price for a thirty-six months period. Average auction price is the average of the three auction prices applying to that year. Average day-ahead price for each zone during the delivery period is the daily average of the corresponding hourly prices. Costs include capacity, transmission, and ancillary services. The ex-post forward premium is computed subtracting from the average of auction prices applying to that year (three auction prices) the costs and the average day-ahead price. The percentage ex-post forward premium is the ex-post forward premium divided by the average auction price. The FMPI is a strip of on-peak monthly futures matching the three-year service period[26] . We compute the FMPI premium as the auction BGS-FP price in each year minus costs minus FMPI. The percentage FMPI premium is the FMPI premium divided by the BGS-FP auction price.

The ex-post forward premium is always positive and ranges from 4.88% to 57.92%. Winning bidders in PJM-BGS are taking price and volumetric risk by selling forward contracts over an amount of electricity that cannot be known with certainty. Winning bidders can satisfy their commitments through self-generation, spot markets, and power derivatives contracts. The first two options are costly and risky. If winning bidders can offset their exposure in the power derivatives market, a positive risk premium is hard to justify in risk management terms. However, hedging price and volumetric risk is problematic. Electricity markets are incomplete. Therefore, quantity risk cannot be perfectly hedged. A precise forecasting of hourly demand during a three-year period is needed. Although load forecasting models can be accurate (e.g. Nedellec et al., 2014), forecasting errors may appear and can be costly. Relying in hedging strategies presents challenges. Woo et al. (2004) suggest a hedging strategy based on standard forwards, and Oum and Oren (2010) posit an approach based on a portfolio of forwards, riskless bonds, and call and put options with various strike prices. Hedging strategies are subject to frictions in their practical implementation (e.g. lack of liquidity of long-term contracts and out-of-the-money options, basis risk in rolling over short-term contracts) and so winning bidders may want a risk premium as compensation for these shortcomings. This demand of a risk premium suggests auction prices (minus costs) may be set above expected spot prices.

---

[26] The PJM-BGS FMPI is based on a strip of futures contracts corresponding to PJM Western Hub Day-Ahead Peak Calendar-Month 5 MW (J4) provided by CME Group. The FMPI contains thirty-six monthly contracts with delivery period between June 1st (year t) and May 31th (year t+3). We define the price of the FMPI as the weighted average of the appropriate thirty-six futures contracts. Weights depend on the present value of each contract. Consider a discount function $w(j)$, with components $w(j) = (1 + r)^{-j/12}$. We define the weight function as $g(t,j) = \frac{w(j)}{\sum_j w(j)}$, indexed by set $j=\{1,2,…,36\}$, corresponding to the price of monthly contracts $Mj(t)$ included in the strip. The weight function integrates to one. Therefore, $FMPI(t) = \sum_j g(t,j)Mj(t)$.

Yearly averages are 34.66% (PSEG), 35.16% (JCPL), 38.07% (ACE), and 42.17% (RECO). Overall, the yearly average is 32.21$/MWh or 38% of the average price of PJM-BGS auctions. As it was the case with CESUR, winning bidders got a positive yearly premium. The size of the premium is 5.26 times higher than in CESUR. The FMPI premium is also positive and ranges from 11.18% to 36%. Yearly averages are 22.18% (PSEG), 21.02% (JCPL), 23.59% (ACE), and 28.29% (RECO). Overall, the yearly average is 23.27$/MWh or 24% of the price of PJM-BGS auctions. FMPI premium is lower than ex-post forward premium. The FMPI premium in PJM-BGS is twenty-two times higher than in CESUR. We stress that PJM-BGS and CESUR offer different contracts. The main difference is that, while CESUR contracts specify the total energy to be provided, PJM-BGS contracts specify the percent of demand that the supplier must satisfy during the delivery period. The amount of demand is unknown at the time they sign the contract. As in CESUR, winning bidders have at their disposal financial instruments to hedge price risk. However, hedging volumetric risk is more complicated. A WB may buy its total electricity need for the three-year period via spot-market purchases. This strategy of reliance on spot markets entails significant risks as exemplified by the April 2001 bankruptcy of Pacific Gas and Electric Company (PG&E), one of the largest utilities in the United States. Alternatively, a WB may buy a strip of futures contracts matching the delivery period. The obstacle here is the computation of the precise number of contracts to buy. In other words, to decide what to buy the WB should solve a risk-constrained expected-cost minimization problem, see Woo et al., (2004) and Kettunen et al. (2010). Solving this problem requires knowing the joint distribution of load and spot price, forward premiums for all futures contracts, the risk preferences of the retailer and an estimate of the risk tolerance of final consumers. Additionally, optimal hedging portfolios may imply complex intra-day trading adjustments, Boroumand et al. (2015). The extent to which the investments needed, and the risk assumed to overcome that obstacle justifies a yearly average premium between 24% and 38% of the average price of PJM-BGS auctions deserves further research.

In summary, we document average ex-post yearly premiums over spot prices of 7.22% (CESUR) and 38% (PJM-BGS) got by WB (suppliers to POLRs). The results when computing premiums using FMPI instead of spot prices are 1.08% (CESUR) and 24% (PJM-BGS). Those facts are consistent with a situation in which sellers want a compensation for assuming price risk (CESUR) and volumetric and price risks (PJM-BGS) and cannot hedge risk by using hedging instruments.

Risk premiums are related to price risk and/or demand volatility in the spot market. Price risk manifests in high volatility[27] and heavy right tail of the price distribution. To gauge whether price risk in the spot market may explain those risk premiums, in Appendix A we present present histograms and summary statistics for daily electricity spot prices in Spain (OMEL market) and in PJM's four zones (ACE, RECO, JCPL, PSEG) in the sample from January 1, 2007, until December 31, 2013. OMEL prices follow a symmetrical distribution with a kurtosis coefficient of 3.81 but PJM prices present skewness coefficients near two, and a kurtosis coefficient near eight. The coefficient of variation is 0.29 with OMEL, and 0.44 with PJM. All these facts suggest that PJM prices are more volatile than OMEL's, and present high prices often. Measures of tail risk are also higher in PJM than in OMEL, as shown in Appendix A

We now turn to the question of what explanatory variables can be related to those ex-post forward premiums. First, we consider variables related to auctions processes. Key variables are the number of allowed bidders and the number of bidding winners. A high number of competitors and winners would hint at low premium, because auction prices should be alike to (expected) future spot prices, because of stiffer competition. As a measure of spot price change, we also include the volatility of daily spot prices three years before each auction. The panel regression model is

$$Ex - Post\ Premium_{jt} = \beta_{j,0} + \sum_{k=1}^{K} \beta_{j,k} X_{j,k,t} + u_{jt} \qquad (3)$$

Where the dependent variable is the standardized yearly ex-post forward premium in monetary units, $X_{j,k,t}$ is a set of $K$ explanatory variables in zone $j$ including the number of start bidders (*Startbidders*), winning bidders (*Wbidders*), and the standard deviation of daily day-ahead electricity prices during the three years before the auction date (*Vol3y*). We include fixed effects by period and estimate parameters in (3) under two choices for *j*, the number of zones. When *j*=1 we consider observations of PJM-BGS. When *j*=2 we consider observations of PJM-BGS and CESUR put together. In estimating (3) we apply a pooled least squared regression with White robust covariances, clustered by cross-section and robust to heteroskedasticity and within cross-section serial correlation. Summary statistics of the control variables are in Table 3.

---

[27] For reasons affecting spot price volatility related with start-up costs, see Reguant (2014).

[INSERT TABLE 3 HERE]

Table 4 reports results of the panel regression model[28] (3).

[INSERT TABLE 4 HERE]

Variable *Startbidders* is significant at conventional levels and presents expected sign in all specifications. An increase in this variable implies a cut in the premium. Variable *Vol3y* is significant and negative suggesting that increases in spot price volatility decrease forward premium[29]. Variable *Wbidders* is non-significant in all cases. These findings deserve two comments. First, the literature on auctions (Kemplerer, 2004), considers the number and characteristics of bidders as important determinants of competition level and results agree with this view. Second, *Vol3y* is significant, even when adding auction-related variables into the model. This suggests all market-related information is not included into auction-related variables[30].

---

[28] Using the FMPI Premium as dependent variable in equation (3) gives similar results. Details are in Appendix A.
[29] This result agrees with Besseminder and Lemmon (2002).

Another noteworthy point is the extent to which market size could explain observed premium. Although CESUR serves a single market, PJM-BGS caters to four zones with different market size, because PSEG and JCPL have more customers than ACE or RECO. As a measure of market size, we use Peak Load Share (PLS) as the percent of total load share auctioned in PJM-BGS in the four areas. On average during 2007-2013 PLS is 2.48% in RECO, 13.11% in ACE, 32.37% in JCPL and 52.04% in PSEG. In the same period, EPFP is highest in RECO (38.52), followed by ACE (32.24), JCPL (28.79) and PSEG (28.48). Therefore, the zone with lowest (highest) market size presents the highest (lowest) ex-post forward premium. However, tests of equality of means fail to reject the null hypotheses of equality of EPFP in the four zones at conventional significance levels. About FMPI premium, average figures suggest that FMPI premium is highest in RECO (29.38), followed by ACE (22.94), PSEG (21.07) and JCPL (19.67). Again, the zone with lowest market size presents the highest ex-post forward premium, but, in contrast with the case with EFPF, the lowest FMPI premium does not belong to the zone with highest market share (PSEG), but to JCPL which has a lower market share. Here, the null hypothesis of equality is rejected because the mean of RECO's FMPI premium is higher than the mean of the other three zones. Next, we analyse the extent to which market size, as measured by PLS, helps in explaining premiums. We include this variable in the panel regression model (3) for ex-post forward premium (EPFP) and FMPI premium. Results suggest that PLS does not contain incremental explanatory power not already included in the number of bidders and the volatility of market prices. Details are in Appendix A.

---

[30] An interesting question is whether the degree of competition is a problem of the auction (design or product offered) or is a market structure problem. If the reason is the former, a policy implication for the market administrators is to look for contracts attractive to prospective bidders and offering alternative products fitted to the characteristics of the suppliers. However, if the reason is a market structure problem, actions should be taken to open the market to new entrants, by lowering barriers to entry. As Fabra and Fabra (2012) point out, with CESUR this low degree of competition is unlikely to be an auction design problem because auction prices converge to forward prices in OTC markets and organized exchanges. Therefore, it looks like a market structure problem based on incentives to raise prices by integrated firms enjoying market power. With PJM-BGS, we do not know published papers addressing this issue.

Market administrators may consider other alternatives to DSA such as spot prices or prices of derivatives contracts. In Appendix A, we present an empirical simulation in the period 2007-2013 suggesting that consumer's aversion to price volatility plays a key role when choosing between alternative methods for the computation of the variable factor of the cost of energy.

### 5.2. Trading activity in power derivatives markets

In this section, we analyze whether market data is consistent with the hypothesis DSA gives incentives to agents to engage in trading activities in power derivatives markets. We analyze volume, open interest, R1 and R2 measures, around auction dates. We compare values of these variables near auction dates with average values in periods without auctions. Next, we test whether significant differences between these measures in normal times and in days surrounding auctions. The estimation window is the first sample available for each price or measure, apart from days surrounding auctions[31]. In doing so, the idea is to capture the 'normal' or "baseline" behavior of each variable or R measure. As for the event window, MacKinlay (1997) suggests using (-1, +1), but other windows are common. Miyajima and Yafeh (2007) suggest (-5, +5) to consider learning and information diffusion. In this paper, we consider five trading days before auctions and five trading days after the auction, the conservative (-5, +5) interval.

---

[31] The initial simple size is N and its average is M(N). We consider an event window of size N2 and the corresponding average is M(N2). Therefore, the sample excluding data within event window has size N1 = N – N2. Thus, its average is M(N1) = (N/N1)*M(N) – (N2/N1)*M(N2).

### 5.2.1. CESUR

Dataset comprises daily data from January 1, 2007, until December 31, 2013, on daily trading volume and open interest for baseload futures contracts traded in OMIP (The Iberian Energy Derivatives Exchange): Yearly, Quarterly, and Monthly. OMIP has provided the data. We use OMIP's FTB (M, Q, Y) base-load futures, with daily settlement (Exchange codes FTBM, FTBQ, FTBYR). As an example, the 1MW baseload Jan13 contract is a monthly swap contract that gives the holder the obligation to buy 1MWh of energy for each hour of January 2013, paying the futures price in Euros/MWh. The seller provides the buyer $1MW \times 24h \times 31days$. The settlement is financial.

We choose liquid contracts within each market segment which are the closest to maturity ones. Within each market segment, we choose three monthly contracts, three quarterly contracts, and two yearly contracts. These contracts representing 99% (99%), 99% (99%) and 100% (100%) of total trading volume (open interest) with monthly, quarterly and yearly contracts, respectively. Monthly, quarterly and yearly contracts account, on average during the sample, for around 95% of total volume, since shorter-term (less than one month) contracts account for around 5% of total volume. Studies on the evolution of trading in this market are Capitán Herráiz, and Rodríguez Monroy (2012), Furió and Meneu (2010) and Capitán Herráiz (2014). Peña and Rodriguez (2016) study time-zero efficiency of European power derivatives markets, including Spain, assessing liquidity and representativeness of organized markets (OMP) versus OTC markets[32].

#### 5.2.1.1. Volume and Open interest

This section summarizes stylized facts on trading volume and open interest. In Appendix A has data on average trading volume, open interest, and summary statistics. Liquid contracts are M1, M2, and Q1, followed by Q2 and Y1, M2, Q3, and Y2. Regarding volatility in trading volume, values of the coefficient of variation suggest a monotonic increase in volatility from M1 (2.10) to Y2 (15.02). All series present right asymmetry and high kurtosis and so Jarque-Bera tests reject the null hypothesis of normality in all contracts. Open interest series present a similar profile to volume series as expected. The Highest liquidity corresponds to contracts M1, M2 and Q1 and volatility follows a similar pattern as in volume. Jarque-Bera tests reject the null hypothesis of normality in all contracts.

---

[32] In the Spanish market, OMP's average market share is 15% in the period 2007-2013.

We turn to the analysis of the behavior of trading volume and open interest of derivatives contracts equivalent to contracts auctioned in CESUR. We concentrate on Q1 contract because a strategy based on the equivalent portfolio made up using monthly contracts might prove difficult to carry out because of low liquidity of contract M3. Figure 1 presents t-statistics for trading volume of Q1 contract in days surrounding CESUR auctions. The null hypothesis is that the average of these days and the average of the full sample (but apart from the auction days) are equal. As we may see, data reject the hypothesis at 1% level on the day of the auction and at 5% level in the day before the auction. Therefore, trading volume increased the day before and the day of CESUR auctions.

[INSERT FIGURE 1 HERE]

Figure 2 presents t-statistics for the open interest of Q1 contract in days surrounding CESUR auctions. The null hypothesis is the average of these days and the average of the full sample (but apart from these days) are equal. As we may see, data reject the hypothesis at 1% level on the day of the auction and at 5% level in the day before the auction. In summary, open interest increased the day before and the day of CESUR auctions.

[INSERT FIGURE 2 HERE]

### 5.2.1.2. Measures of speculation and hedging

Results suggest increases in trading activity near CESUR auctions. We are interested in knowing whether hedging or speculative strategies drive market's activity. Next, we present stylized facts on measures of speculation and hedging R1 and R2[33]. With R1, average values vary from M3 (0.104) to Y1 (0.010). The higher the average value, the higher is the speculative trading in a contract. Thus, this measure suggests contracts M3, M2, and M1are more likely to be used for speculative trading. Volatilities present a range of variation going from M3 (0.557) to Q1 (0.036). High volatility in a contract suggests wide variations between speculative and hedging situations. With R2, average values vary from M1 (1.84) to M3 (0.56). The higher the average value, the higher is the speculative trading in a contract and so this measure suggest contracts M1, M2, Q1, Q2, and Y1 are more likely to be used for speculative trading. Volatilities present a range of change going from M1 (4.25) to Y2 (1.015). Median values are higher for M2, M1, Q2, Q1 and Y1 suggesting again these contracts are more likely to be used in speculative trading. Summarizing the results, measure R1 suggests contracts M3, M2, and M1 are more likely to be used for speculative trading, but R2 suggests M1, M2, Q1, Q2, and Y1. We turn to the analysis of measures R1 and R2 in dates surrounding each CESUR auction. Table A2.5 in the Appendix A summarizes t-statistics for contracts and measures. We compare the mean of days surrounding CESUR auctions against the mean of the rest of the sample. If t-stats are positive (negative) and significant, the corresponding measure (R1 or R2) is higher (lower) during this day than in the rest of the sample. There are 94 significant t-stats out of 176; 91 of them are negative and 3 are positive. Decreases (increases) in R measures are associated with increases in hedging (speculative) activity. Therefore, the bulk of evidence suggests hedging increased in dates surrounding CESUR auctions.

---

[33] To save space we summarize results. Detailed tables and graphs are available on Appendix A.

### 5.2.2. PJM-BGS

CME provides futures prices for three different hubs: Dayton Hub, N. Illinois Hub and Western Hub. After analyzing the relative liquidity of contracting in each hub, we choose Western Hub as the most liquid and representative. We consider off-peak and on-peak contracts. The most liquid off-peak contract is the PJM Western Hub Day-Ahead Off–Peak Calendar Month 5 MW (E4) and the reference for on-peak prices is the PJM Western Hub Day-Ahead Peak Calendar-Month 5 MW (J4). We analyze thirty-six monthly contracts of both references (from M4 to M39) because they span the three-year auction's delivery period[34]. As an illustration, we concentrate on the most liquid contract, the E4 front contract E4M4[35]. In Figure 3 we present t-statistics for trading volume of E4M4 contract in days surrounding auctions. The null hypothesis is that the average of those days and the average of the full sample (except auction days) are equal. Data rejects the null hypothesis at 1% level on the day of the auction and at 5% level in the day before the auction. So, trading volume increased on the day before and on the day of the initiation of PJM-BGS auctions.

[INSERT FIGURE 3 HERE]

Figure 4 presents t-statistics for the open interest of E4M4 contract in days surrounding auctions. The null hypothesis is the average of these days and the average of the full sample (apart from auction days) is equal. Data do not reject the null hypothesis at conventional

---

[34] Details and summary statistics are in Appendix A.
[35] This is the contract with highest liquidity. Results for other contracts are in Appendix A.

significance levels. In summary, open interest did not increase the days surrounding auctions.

[INSERT FIGURE 4 HERE]

Therefore, empirical evidence supports the hypothesis (2) to this extent because trading activity increases volume but not open interest. We turn to the analysis of measures R1 and R2 in dates surrounding auctions. In Appendix A, Table A2.6 summarizes t-statistics for all liquid contracts. We compare the mean of days surrounding PJM-BGS auctions against the mean of days of the rest of the sample. If the t-stat is positive (negative) and significant, the corresponding measure (R1 or R2) is higher (lower) during this day than in the rest of the sample. We report 880 t-statistics of which 272 are significant and positive and 12 significant and negative. Increases in R measures are associated with increases in speculation. So, empirical evidence suggests speculative activity increased in dates surrounding auctions. This result is consistent with the evidence on trading volume and open interest.

## 6. Conclusions and Policy Implications

Empirical evidence on the consequences of default supply auctions in electricity markets is far from decisive. If auctions are not competitive enough and winning bidders have market power, the specific conditions of the electricity market favour winning bidders. Small firms and households have an inelastic demand and will pay high prices rather than be cut off. On the other hand, if winning bidders assume volumetric risk, as in PJM-BGS, the difficulties of hedging this risk may imply incurring losses that should be rewarded. This paper contributes to the literature on the results of default supply auctions by studying ex-post forward premium and forward market price index premium got by winning bidders. We also examine speculation and hedging activities in power derivatives markets in dates near auctions.

Based on an extensive database in Spain (CESUR auctions) and in New Jersey (PJM-BGS auctions), we document the following facts. First, in the period 2007-2013 winning bidders in CESUR and PJM-BGS got an average ex-post yearly forward premium of 7.22% (CESUR) and 38% (PJM-BGS). The premium using an index of futures prices is 1.08% (CESUR) and 24% (PJM-BGS). Participants in the CESUR auction know the amount of electricity to be delivered and winning bidders have at their disposal financial instruments to hedge price risk. In PJM-BGS, they do not know the amount of electricity, and available financial contracts allow for the hedging of price risk. Hedging volumetric risk is complex and therefore further research is needed to understand the reasons justifying observed premium.

Second, empirical evidence suggests a negative relationship between the number of bidders and the ex-post forward premium. Spot price uncertainty, as measured by the spot market price volatility, also has a negative effect on the ex-post forward premium. Third, trading in power derivatives markets increased in days surrounding auctions. Fourth, with CESUR hedging-driven strategies predominate around auction dates, but with PJM-BG, speculation-driven trading prevails.

Market administrators may consider other alternatives to DSA such as spot prices or prices of derivatives contracts. An empirical simulation in the period 2007-2013 suggests that consumer's aversion to price volatility plays a key role when choosing between alternative methods for the computation of the variable factor of the cost of energy. Consequently, our policy recommendation to market regulators and administrators is that they should gauge consumers' price risk aversion before introducing alternative methods to DSA for the computation of the cost of energy part included in the electricity bill. The impact of price risk aversion on consumer choice documented here provides an added insight into the results in Holland and Mansur (2006) in the sense that flat rate change by month captures a significant part of the efficiency gains of time-varying prices.

While several techniques in this paper have provided useful insights, the range of contracts used in the analysis are limited. Shorter-term contracts should yield a fuller perspective on hedging versus speculative activities. Understanding the impact of volumetric risk on premium is another promising area of research. Looking forward, application of methods in this paper to other default supply auctions, and power derivatives markets are immediate extensions, which we left for future research.

**References**

ap Gwilym, O., Buckle, M. and Evans, P., 2002. The volume-maturity relationship for stock index, interest rate and bond futures contracts. EBMS Working Paper EBMS/2002/3.

Bessembinder, H. and Seguin, P. J., 1993. Price volatility, trading volume, and market depth: evidence from futures markets. Journal of Financial and Quantitative Analysis, 28, 21–39.

Bessembinder, H. and Lemmon, M. L., 2002. Equilibrium pricing and Optimal Hedging in Electricity Forward Markets. Journal of Finance, 57, 1347-1382.

Boroumand, R. H., Goutte, S., Porcher, S., and Porcher, T., 2015. Hedging strategies in energy markets: the case of electricity retailers. Energy Economics, 51, 503–509.

Capitán Herráiz, A., 2014. Regulatory proposals for the development of an efficient Iberian energy forward market. Ph.D. Dissertation. ETSII. Universidad Politécnica de Madrid.

Capitán Herráiz, A. and C. Rodríguez Monroy, 2009. Evaluation of the Liquidity in the Iberian Power Futures Market. IV Congress of Spanish Association of Energy Economists (AEEE), Seville (Spain), January 2009.

Capitán Herráiz, A. and C. Rodríguez Monroy, 2012. Evaluation of the trading development in the Iberian Energy Derivatives Market. Energy Policy, 51, 973-984.

Cartea, A. and P. Villaplana, 2008. Spot price modeling and the valuation of electricity forward contracts: the role of demand and capacity. Journal of Banking and Finance, 32, 2502-2519.

Cartea, A. and P. Villaplana, 2012. Analysis of the main determinants of electricity forward prices and forward risk premia, in Quantitative energy finance: modeling, pricing and hedging in energy and commodity markets, 1-29. Springer-Verlag.

CNE, 2008. Informe de valoración preliminar sobre las subastas de emisión primarias de energía y CESUR. http://www.cne.es/

CNMC, 2014. Informe sobre el desarrollo de la 25ª subasta CESUR previsto en el artículo 14.3 de la orden ITC/1659/2009, de 22 de Junio. www.cnmc.es.

de Castro, L., Negrete-Pincetic, M., Gross, G., 2008. Product Definition for Future Electricity Supply Auctions: The 2006 Illinois Experience. The Electricity Journal 21(7): 50–62.

De Frutos, M.A. and Fabra, N. , 2012. How to Allocate Forward Contracts: the case of electricity markets. European Economic Review, 56(3), 451-469.

Ederington, L. and Lee, J. H., 2002. Who trades futures and how: evidence from the heating oil futures market, Journal of Business, 75, 353–73.

Federico, G., Vives, X., 2008. Competition and regulation in the Spanish gas and electricity Markets. Reports of the Public-Private Sector Research Center, 1, IESE Business School, University of Navarra.

Fabra, N. and J. Fabra Utray, 2012. El Déficit Tarifario en el Sector Eléctrico Español. Papeles de Economía Española 134, 2-14.

Furió, D. and V. Meneu, 2010. Expectations and forward risk premium in the Spanish deregulated power market. Energy Policy, 38, 784-793.

García, P., Leuthold, R. M. and Zapata, H. ,1986. Leadlag relationships between trading volume and price variability: new evidence, The Journal of Futures Markets, 6, 1–10.

Glachant, J-M, D. Finon and A. de Hauteclocque, 2011. Competition, Contracts, and Electricity Markets. Edward Elgar Publishing.

Haugom, E. and C.J. Ullrich, 2012. Market efficiency and risk premia in short-term forward prices. Energy Economics, 34, 1931-1942.

Holland, S.P. and E. T. Mansur, 2006. The Short-Run Effects of Time-Varying Prices in Competitive Electricity Markets. The Energy Journal, 27, 4, 127-155.

Ito, K. and M. Reguant, 2016. Sequential Markets, Market Power, and Arbitrage. American Economic Review, 106, 1921–1957.

Kemplerer, P., 2004. <u>Auctions: Theory and Practice</u>. Princeton University Press.

Kettunen, J., Salo, A., and Bunn, D. W., 2010. Optimization of electricity retailer's contract portfolio subject to risk preferences. <u>IEEE Transactions on Power Systems</u>, 25, 1, 117–128.

Lacasse, Ch. And T. Wininger, 2007. Maryland vs. New Jersey: Is there a "best" competitive bid process?". <u>The Electricity Journal</u>, 20, 3, 46-59.

Leuthold, R. M., 1983. Commercial use and speculative measures of the livestock commodity futures markets, <u>The Journal of Futures Markets</u>, 3, 113–35.

Loxley, C. and D. Salant, 2004. Default Service Auctions. <u>Journal of Regulatory Economics</u>, 27, 3, 201-229.

Lucia, J., Pardo, Á., 2010. On measuring speculative and hedging activities in futures markets from volume and open interest data. <u>Applied Economics</u>, 42, 1549-1557.

Maurer, L.T.A. and L.A. Barroso, 2011. <u>Electricity Auctions: An Overview of Efficient Practices</u>. The World Bank.

MacKinlay, A. C., 1997. Event studies in economics and finance. <u>Journal of Economic Literature</u>, 35, 13–39.

Miyajima, H. and Yafeh, Y., 2007. Japan's banking crisis: an event-study perspective. <u>Journal of Banking & Finance</u> 31, 2866-2885.

Nedelle, R., J. Cugliari, and Y. Goude, 2014. GEFCom2012: Electric load forecasting and backcasting with semi-parametric models. <u>International Journal of Forecasting</u>, 30, 375-381.

Negrete-Pincetic, M., L. de Castro and H.A. Pulgar-Painemal, 2015. Electricity supply auctions: Understanding the consequences of the product definition. <u>Electrical Power and Energy Systems</u>, 64, 285–292

Newbery, D.M. and T. McDaniel, 2003. Auctions and Trading In Energy Markets - An Economic Analysis. Ch 10, pp 95-234 in CRI Regulatory Review 2002/ 2003 P. Vass (ed.) Bath: Centre for the Study of Regulated Industries.

Oum, Y., and S. S. Oren, 2010. Optimal Static Hedging of Volumetric Risk in a Competitive Wholesale Electricity Market. <u>Decision Analysis</u>, 7,1, 107-122.

Peña, J. I., and R. Rodriguez, 2016. Time-Zero Efficiency of European Power Derivatives Markets. <u>Energy Policy</u>, 95, 253-268.

Redl, C., R. Haas, C. Huber and B. Bohm, 2009. Price formation in electricity forward markets and the relevance of systematic forecast errors. <u>Energy Economics</u>, 31, 356-364.

Reguant, M., 2014. Complementary Bidding Mechanisms and Startup Costs in Electricity Markets. <u>Review of Economic Studies</u>, 81, 1708-1742.

Robinson, D. ,2015. The Scissors Effect. Oxford Institute for Energy Studies. Paper EL 14.

Woo, C-K, R. I. Karimov and I. Horowitz, 2004. Managing electricity procurement cost and risk by a local distribution company. <u>Energy Policy</u>, 32, 635-645.

# Table 1: CESUR auctions (baseload products)

| Auction Number | Date | Product | CESUR Price (€/MWh) | Average spot price during delivery period SP (€/MWh) | Ex-post forward premium (€/MWh) | % Ex-post forward premium | FMPI (€/MWh) | Premium FMPI (€/MWh) | % FMPI premium |
|---|---|---|---|---|---|---|---|---|---|
| 1 | 19/06/2007 | Q3-07 | 46.27 | 36.45 | 9.82 | 21.22% | 44.45 | 1.82 | 3.93% |
| 2 | 18/09/2007 | Q4-07 | 38.45 | 47.78 | -9.33 | -24.27% | 38.58 | -0.13 | -0.34% |
| 3 | 18/12/2007 | Q1-08 | 64.65 | 65.85 | -1.2 | -1.86% | 64.2 | 0.45 | 0.70% |
| **Average** | | | **49.79** | **50.03** | **-0.24** | **-1.63%** | | **0.71** | **1.43%** |
| 4(1) | 13/03/2008 | Q2-08 | 63.36 | 56.92 | 6.44 | 10.16% | 61.8 | 1.56 | 2.46% |
| 4(2) | 13/03/2008 | Q2Q3-08 | 63.73 | 63.7 | 0.03 | 0.05% | 61.8 | 1.93 | 3.03% |
| 5(1) | 17/06/2008 | Q3-08 | 65.15 | 70.41 | -5.26 | -8.07% | 64.5 | 0.65 | 1.00% |
| 5(2) | 17/06/2008 | Q3Q4-08 | 65.79 | 67.53 | -1.74 | -2.64% | 64.5 | 1.29 | 1.96% |
| 6 | 25/09/2008 | Q4-08 | 72.49 | 64.65 | 7.84 | 10.82% | 72.55 | -0.06 | -0.08% |
| 7 | 16/12/2008 | Q1-09 | 58.86 | 43.1 | 15.76 | 26.78% | 58.55 | 0.31 | 0.53% |
| **Average** | | | **64.90** | **61.05** | **3.85** | **6.18%** | | **0.95** | **1.48%** |
| 8 | 26/03/2009 | Q2-09 | 36.58 | 36.99 | -0.41 | -1.12% | 36.1 | 0.48 | 1.31% |
| 9(1) | 25/06/2009 | Q3-09 | 42 | 35.05 | 6.95 | 16.55% | 41.25 | 0.75 | 1.79% |
| 9(2) | 25/06/2009 | Q4-09 | 45.67 | 32.87 | 12.8 | 28.03% | 44.25 | 1.42 | 3.11% |
| 10(1) | 15/12/2009 | Q1-10 | 39.43 | 25.38 | 14.05 | 35.63% | 39.1 | 0.33 | 0.84% |
| 10(2) | 15/12/2009 | Q2-10 | 40.49 | 34.97 | 5.52 | 13.63% | 39.1 | 1.39 | 3.43% |
| **Average** | | | **40.83** | **33.05** | **7.78** | **18.54%** | | **0.87** | **2.10%** |
| 11 | 23/06/2010 | Q3-10 | 44.5 | 44.07 | 0.43 | 0.97% | 44.05 | 0.45 | 1.01% |
| 12 | 21/09/2010 | Q4-10 | 46.94 | 43.33 | 3.61 | 7.69% | 46.83 | 0.11 | 0.23% |
| 13 | 14/12/2010 | Q1-11 | 49.07 | 45.22 | 3.85 | 7.85% | 49.15 | -0.08 | -0.16% |
| **Average** | | | **46.84** | **44.21** | **2.63** | **5.50%** | | **0.16** | **0.36%** |
| 14 | 22/03/2011 | Q2-11 | 51.79 | 48.12 | 3.67 | 7.09% | 51.35 | 0.44 | 0.85% |
| 15 | 28/06/2011 | Q3-11 | 53.2 | 54.23 | -1.03 | -1.94% | 52.9 | 0.3 | 0.56% |
| 16 | 27/09/2011 | Q4-11 | 57.99 | 52.01 | 5.98 | 10.31% | 57.95 | 0.04 | 0.07% |
| 17 | 20/12/2011 | Q1-12 | 52.99 | 50.64 | 2.35 | 4.43% | 52.75 | 0.24 | 0.45% |
| **Average** | | | **53.99** | **51.25** | **2.74** | **4.97%** | | **0.26** | **0.48%** |
| 18 | 21/03/2012 | Q2-12 | 51 | 46.07 | 4.93 | 9.67% | 50.75 | 0.25 | 0.49% |
| 19 | 26/06/2012 | Q3-12 | 56.25 | 49.09 | 7.16 | 12.73% | 53.05 | 3.2 | 5.69% |
| 20 | 25/09/2012 | Q4-12 | 49.25 | 43.16 | 6.09 | 12.37% | 49.34 | -0.09 | -0.18% |
| 21 | 21/12/2012 | Q1-13 | 54.18 | 40.34 | 13.84 | 25.54% | 54.07 | 0.11 | 0.20% |
| **Average** | | | **52.67** | **44.67** | **8.01** | **15.08%** | | **0.87** | **1.55%** |
| 22 | 20/03/2013 | Q2-13 | 45.41 | 34.26 | 11.15 | 24.55% | 45.25 | 0.16 | 0.35% |
| 23 | 25/06/2013 | Q3-13 | 47.95 | 49.81 | -1.86 | -3.88% | 47.9 | 0.05 | 0.10% |
| 24 | 24/09/2013 | Q4-13 | 47.58 | 54.73 | -7.15 | -15.03% | 47.55 | 0.03 | 0.06% |
| **Average** | | | **46.98** | **46.27** | **0.71** | **1.88%** | | **0.08** | **0.17%** |
| **Averages** | | | **50.86** | **47.22** | **3.64** | **7.22%** | | **0.56** | **1.08%** |

# Table 2: PJM-BGS auctions

This table reports basic facts on the 28 PJM-BGS auctions. Columns contain auction year, POLR, Auction BGS-FP price (the auction price for a thirty-six months period), Average Auction Price (the average of the three auction prices applying to that year), Average Day-Ahead Price during delivery period, Costs, Ex-post forward premium (Average Auction Price – Costs – Average Day-ahead price during delivery period), and % Ex-post forward premium (Ex-post forward premium/Average Auction Price), FMPI , FMPI premium (Auction BGS-FP price – costs - FMPI) and FMPI Premium % (FMPI premium/Auction BGS-FP price).

| YEAR | POLR | Auction BGS-FP Price $/MWH | Average Auction Price $/MWH | Average Day-Ahead Price during delivery period $/MWH | Costs $/MWH | Ex-post Forward Premium $/MWH | Ex-post Forward Premium % | FMPI $/MWH | FMPI Premium $/MWH | FMPI Premium % |
|---|---|---|---|---|---|---|---|---|---|---|
| 2007 | ACE | 99.59 | 90.02 | 70.79 | 11.34 | 7.89 | 10.02% | 72.01 | 16.24 | 16.31% |
| 2008 | ACE | 116.5 | 106.69 | 66.66 | 15.01 | 25.02 | 27.29% | 84.02 | 17.47 | 15.00% |
| 2009 | ACE | 105.36 | 107.15 | 41.29 | 17.44 | 48.42 | 53.97% | 65.97 | 21.95 | 20.83% |
| 2010 | ACE | 98.56 | 106.81 | 51.72 | 18.03 | 37.06 | 41.74% | 58.01 | 22.52 | 22.85% |
| 2011 | ACE | 100.95 | 101.62 | 39.55 | 15.73 | 46.34 | 53.96% | 52.65 | 32.57 | 32.26% |
| 2012 | ACE | 85.1 | 94.87 | 37.76 | 14.51 | 42.60 | 53.01% | 44.72 | 25.87 | 30.40% |
| 2013 | ACE | 87.27 | 91.11 | 56.22 | 16.56 | 18.33 | 24.58% | 46.75 | 23.96 | 27.46% |
| **Average** | | | | | | **32.24** | **37.80%** | | **22.94** | **23.59%** |
| 2007 | PSE&G | 98.88 | 88.93 | 72.41 | 11.34 | 5.18 | 6.68% | 72.01 | 15.53 | 15.71% |
| 2008 | PSE&G | 111.5 | 104.3 | 66.57 | 15.01 | 22.72 | 25.44% | 84.02 | 12.47 | 11.18% |
| 2009 | PSE&G | 103.72 | 104.7 | 41.72 | 17.44 | 45.54 | 52.19% | 65.97 | 20.31 | 19.58% |
| 2010 | PSE&G | 95.77 | 103.66 | 52.35 | 18.03 | 33.28 | 38.86% | 58.01 | 19.73 | 20.60% |
| 2011 | PSE&G | 94.3 | 97.93 | 39.66 | 15.73 | 42.54 | 51.75% | 52.65 | 25.92 | 27.49% |
| 2012 | PSE&G | 83.88 | 91.32 | 40.17 | 14.51 | 36.64 | 47.70% | 44.72 | 24.65 | 29.39% |
| 2013 | PSE&G | 92.18 | 90.12 | 60.12 | 16.56 | 13.44 | 18.27% | 46.75 | 28.87 | 31.32% |
| **Average** | | | | | | **28.48** | **34.41%** | | **21.07** | **22.18%** |
| 2007 | JCP&L | 99.64 | 88.59 | 73.48 | 11.34 | 3.77 | 4.88% | 72.01 | 16.29 | 16.35% |
| 2008 | JCP&L | 114.09 | 104.72 | 65.19 | 15.01 | 24.52 | 27.34% | 84.02 | 15.06 | 13.20% |
| 2009 | JCP&L | 103.51 | 105.75 | 41.08 | 17.44 | 47.23 | 53.48% | 65.97 | 20.1 | 19.42% |
| 2010 | JCP&L | 95.17 | 104.26 | 51.66 | 18.03 | 34.57 | 40.09% | 58.01 | 19.13 | 20.10% |
| 2011 | JCP&L | 92.56 | 97.08 | 39.26 | 15.73 | 42.09 | 51.74% | 52.65 | 24.18 | 26.12% |
| 2012 | JCP&L | 81.76 | 89.83 | 38.03 | 14.51 | 37.29 | 49.51% | 44.72 | 22.53 | 27.56% |
| 2013 | JCP&L | 83.7 | 86.01 | 57.40 | 16.56 | 12.05 | 17.36% | 46.75 | 20.39 | 24.36% |
| **Average** | | | | | | **28.79** | **34.91%** | | **19.67** | **21.02%** |
| 2007 | RECO | 109.99 | 97.64 | 71.45 | 11.34 | 14.85 | 17.20% | 72.01 | 26.64 | 24.22% |
| 2008 | RECO | 120.49 | 113.87 | 64.91 | 15.01 | 33.95 | 34.34% | 84.02 | 21.46 | 17.81% |
| 2009 | RECO | 112.7 | 114.39 | 40.80 | 17.44 | 56.15 | 57.91% | 65.97 | 29.29 | 25.99% |
| 2010 | RECO | 103.32 | 112.17 | 50.06 | 18.03 | 44.08 | 46.82% | 58.01 | 27.28 | 26.40% |
| 2011 | RECO | 106.84 | 107.62 | 38.44 | 15.73 | 53.45 | 58.16% | 52.65 | 38.46 | 36.00% |
| 2012 | RECO | 92.51 | 100.89 | 40.58 | 14.51 | 45.80 | 53.03% | 44.72 | 33.28 | 35.97% |
| 2013 | RECO | 92.58 | 97.31 | 59.41 | 16.56 | 21.34 | 26.43% | 46.75 | 29.27 | 31.62% |
| **Average** | | | | | | **38.52** | **41.99%** | | **29.38** | **28.29%** |
| **Total Average** | | | | | | **32.00** | **37.28%** | | **23.27** | **23.77%** |

**Table 3: Summary Statistics Explanatory Variables**

This table reports the summary statistics of explanatory variables in equation (3) The sample period is January 1, 2007 to December 31, 2013. Auction data comes from fifty-six auction prices obtained from twenty-four CESUR auctions with twenty-eight products and prices and seven PJM-BGS auctions in four zones, yielding twenty-eight prices. When aggregating the data in yearly terms, we get seven observations in CESUR and twenty-eight in PJM-BGS, totaling thirty-five prices.

| Variable | Mean | Std. Dev | Min | Max |
|---|---|---|---|---|
| **Joint sample** | | | | |
| *Wbidders* | 18.10 | 4.69 | 9.00 | 23.00 |
| *Startbidders* | 20.83 | 4.92 | 15.00 | 32.00 |
| *Vol3y* | 17.86 | 4.52 | 8.12 | 31.70 |
| **PJM-BGS** | | | | |
| *Wbidders* | 10.00 | 2.26 | 8.00 | 13.00 |
| *Startbidders* | 19.00 | 3.51 | 15.00 | 25.00 |
| *Vol3y* | 22.51 | 5.60 | 18.11 | 31.70 |
| **CESUR** | | | | |
| *Wbidders* | 12.00 | 2.42 | 9.00 | 15.00 |
| *Startbidders* | 28.00 | 3.13 | 24.00 | 32.00 |
| *Vol3y* | 12.29 | 2.32 | 8.12 | 23.34 |

**Table 4: Panel Regression**

This table reports the results of a panel regression in which the dependent variable is the standardized (by markets) yearly ex-post premium. The sample period is January 1, 2007 to December 31, 2013. Auction data comes from fifty-six auction prices obtained from twenty-four CESUR auctions with twenty-eight products and prices and seven PJM-BGS auctions in four zones, yielding twenty-eight prices. When aggregating the data in yearly terms, we get seven observations in CESUR and twenty-eight in PJM-BGS, totaling thirty-five prices. Explanatory variables are: *Wbidders* is the number of winning bidders; *Startbidders*, is the number of authorized bidders; *Vol3y* is spot price volatility three years before the auction. ***, and ** are significant at 1%, and 5%, respectively. Figures in parenthesis are t-statistics.

|  | (1) PJM | (2) PJM+CESUR | (3) PJM | (4) PJM+CESUR |
|---|---|---|---|---|
| *Vol3y* | -0.018** | -0.095*** | -0.017** | -0.064*** |
|  | (-3.58) | (-11.83) | (-5.63) | (-4.84) |
| *Startbidders* | -0.483*** | -0.083*** | -0.483*** | -0.141*** |
|  | (-99.04) | (-16.81) | (-110.59) | (-5.87) |
| *Wbidders* |  |  | -0.0005 | 0.037 |
|  |  |  | (-0.12) | (2.58) |
| *Constant* | 8.439*** | 2.088** | 8.434*** | 2.501*** |
|  | (85.85) | (4.28) | (82.29) | (5.19) |
| Observations | 28 | 35 | 28 | 35 |
| R-squared | 0.997 | 0.869 | 0.997 | 0.893 |
| Adjusted R-squared | 0.995 | 0.829 | 0.995 | 0.855 |
| rss | 0.0826 | 3.926 | 0.0825 | 3.203 |
| rmse | 0.0642 | 0.389 | 0.0659 | 0.358 |
| t statistics in parentheses |  |  |  |  |
| ** p<0.05, *** p<0.01 |  |  |  |  |

**Figure 1: Trading Volume Q1 contract t-statistic around CESUR auctions**

This figure illustrates the evolution of trading volume of Q1 contract. Figure shows t-statistics near dates of CESUR auctions. The null hypothesis is that the average trading volume in these days and the average of the full sample (excluding days near auctions) are equal. Dataset consists of daily data from January 1, 2007 to December 31, 2013.

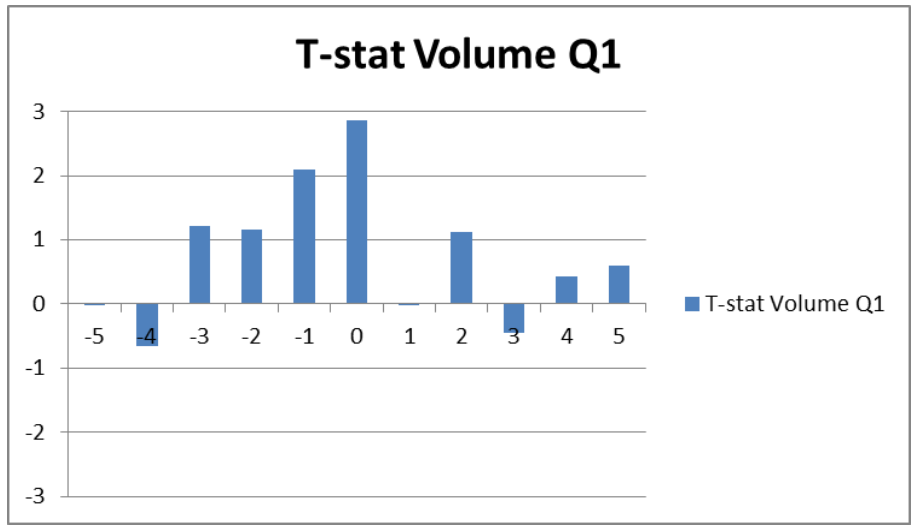

**Figure 2: Open Interest Q1 contract t-statistic around CESUR auctions**

This figure illustrates the evolution of open interest of Q1 contract. Figure shows t-statistics near dates of CESUR auctions. The null hypothesis is that the average open interest in these days and the average of the full sample (excluding days near auctions) are equal. Dataset consists of daily data from January 1, 2007, until December 31, 2013.

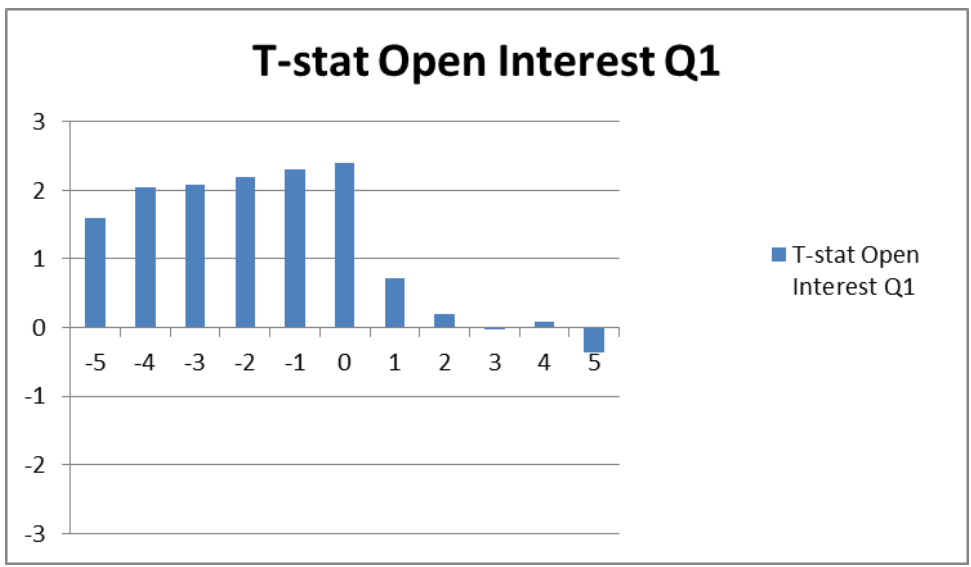

**Figure 3: Trading Volume t-statistic around PJM-BGS auctions, E4M4**

This figure illustrates the evolution of trading volume of E4M4 contract. Figure shows t-statistics near dates of PJM-BGS auctions. The null hypothesis is that the average trading volume in these days and the average of the full sample (excluding days near auctions) are equal. Dataset consists of daily data from January 1, 2007, until December 31, 2013.

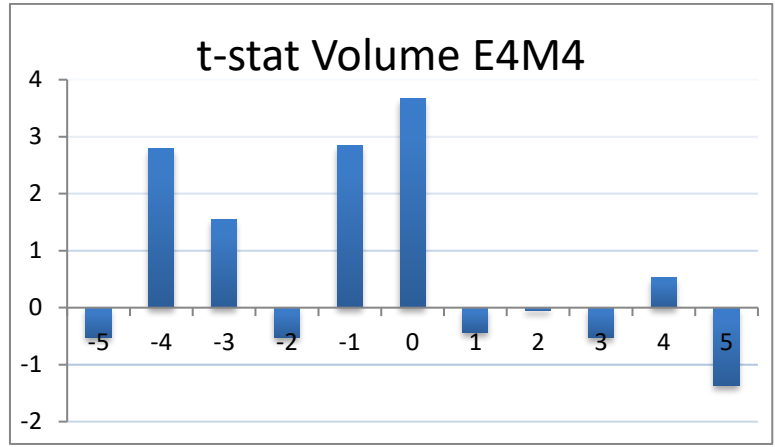

**Figure 4: Open Interest t-statistic around PJM-BGS auctions, E4M4**

This figure illustrates the evolution of open interest of E4M4 contract. Figure shows t-statistics near dates of PJM-BGS auctions. The null hypothesis is that the average open interest in these days and the average of the full sample (excluding days near auctions) are equal. Dataset consists of daily data from January 1, 2007, until December 31, 2013.

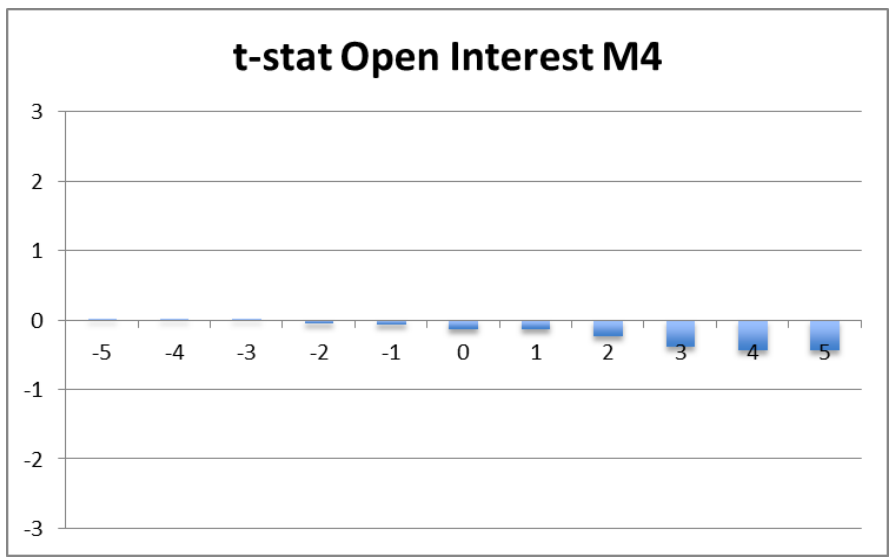